\documentclass[11pt]{article}
\usepackage{amsbsy}
\usepackage{amsfonts}
\usepackage{amsmath}
\usepackage{amsthm}
\usepackage{amssymb}
\usepackage{apacite}
\usepackage{setspace}
\usepackage{graphicx,psfrag,epsf}
\usepackage{bm,multicol}
\usepackage[english]{babel}
\usepackage{natbib}
\usepackage[T1]{fontenc}
\usepackage[utf8]{inputenc}
\usepackage{authblk}
\usepackage{xcolor}
\usepackage{natbib}

\usepackage[title]{appendix}

\newtheorem{lemma}{Lemma}

\newtheorem{theorem}{Theorem}

\newtheorem{definition}{Definition}

\newtheorem{remark}{Remark}

\doublespace
\newcommand{\newc}{\newcommand}
\newc{\N}{\mbox{N}}
\newc{\1}{\bf{1}}
\def\signed #1{{\leavevmode\unskip\nobreak\hfil\penalty50\hskip2em
  \hbox{}\nobreak\hfil(#1)%
  \parfillskip=0pt \finalhyphendemerits=0 \endgraf}}

\newsavebox\mybox

\begin{document}
\title{From $p$-Values to Posterior Probabilities of Hypothesis}
\author[1]{D. Vélez}
\author[2]{ M.E. Pérez}
\author[2]{L. R. Pericchi}
\affil[1]{University of Puerto Rico, Río Piedras Campus, Statistical Institute and Computerized Information Systems, Faculty of Business Administration, 15 AVE Universidad STE 1501, San Juan, PR 00925-2535, USA}
\affil[2]{University of Puerto Rico, Río Piedras Campus, Department of Mathematics, Faculty of Natural Sciences, 17 AVE Universidad STE 1701, San Juan, PR 00925-2537, USA}
\date{}
\maketitle
\begin{abstract}
Minimum Bayes factors are commonly used to transform two-sided p-values to lower bounds on the posterior probability of the null hypothesis, as in \citet{art11}. In this article, we show posterior probabilities of hypothesis by transforming the commonly used $-e\cdot p\cdot \log(p)$, proposed by \citet{art15} and \citet{art4}. This is achieved after adjusting this minimum Bayes factor with the information to approximate it to an exact Bayes factor, not only when $p$ is a $p$-value but also when $p$ is a pseudo $p$-value in the sense of \citet{CasBer2001}. Additionally we show the fit to a refined version to linear models.
\end{abstract}

\newpage
\section{Pseudo $P$-Values}
Under the null hypotheses, $p$-values are well known to have
Uniform(0,1), in \citet{CasBer2001} a more general definition is given

\begin{definition}\label{def1}
A \textbf{$p$-value} $p(\textbf{X})$ is a statistic satisfying $0\leq p(\textbf{x})\leq 1$ for every sample point \textbf{x}. Small values of $p(\textbf{X})$ give evidence that $H_1$ is true. A $p$-value is \textbf{valid} if, for every $\theta\in \Theta_0$ and every $0\leq\alpha\leq 1$,
$$P_\theta(p(\textbf{X})\leq \alpha)\leq \alpha.$$
\end{definition}

\begin{remark}\label{remark1}
We consider any \textbf{$p$-value} complying the Definition~\ref{def1} without equality for all $\alpha$ a \textbf{pseudo $p$-value}.
\end{remark}

The ``Robust Lower Bound" ($RLB$) as is called in \citet{art11} and proposed by \citet{art4} is: \begin{equation*}
B_L(p)=\begin{cases}
-e\cdot p\cdot\log(p)& p<e^{-1}\\
~~~~~~~~~1 & \text{otherwise}
\end{cases}
\end{equation*}

when under $H_0$ $p$ is Uniform(0,1) and the density of $p$ under $H_1$ is $Beta(\xi,1)$ for $0<\xi<1$.
Note that this calibration has been proposed already in \citet{art15}.
Another class of decreasing densities is $Beta(1,\xi)$ with $\xi>1$. This leads to the "$-e\cdot q\cdot\log(q)$" calibration, where $q=1-p$ see \citet{art14}.\\

\begin{figure}[h]
    \centering
     \includegraphics[scale=.3]{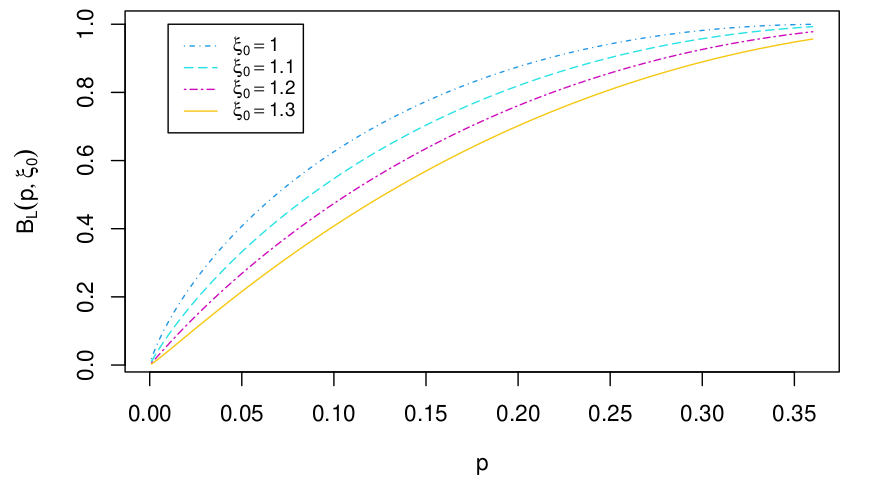}  
    \caption{Graph the $RLB_{\xi_0}$ for different $\xi_0$.}
    \label{fig0}
\end{figure}

In contrast with the Remark~\ref{remark1} if we consider $p(x)$ a pseudo $p$-value under $H_0$, that is, $$p\sim Beta(\xi_0,1)~~~\text{with}~~~\xi_0>1, ~~\text{fixed but arbitrary,}$$
under the test $$H_0:p\sim Beta(\xi_0,1)~~~vs~~~H_1:p\sim f(p|\xi)$$
with $f(p|\xi)\sim Beta(\xi,1)$ for $0<\xi<1$, the $RLB_{\xi_0}$ is 
\begin{equation}\label{eq1}
B_L(p,\xi_0)=\begin{cases}
-e\cdot \xi_0\cdot p^{\xi_0}\log(p)& p<e^{-1}\\
~~~~~~~~~1 & \text{otherwise}
\end{cases}
\end{equation}

where $\xi_0$ has to be estimated or calculated theoretically, but we known that $\xi_0=1$ when the $p$-value is not pseudo $p$-value.\\
On the other hand, since $f(p|\xi)=\xi p^{\xi-1}$ has its maximum in $\xi=-\frac{1}{\log(p)}<1$ with $p<e^{-1}$ then $f(p|\xi)$ is decreasing for $\xi>-\frac{1}{\log(p)}$, thus for any Bayes Factor $B_{01}$

$$B_{01}\geq B_{L}(p) >B_L(p,\xi_0) ~~~\text{con}~~~\xi_0>1$$
see Figure~\ref{fig0}. 

In the sequel, we want to calibrate $RLB_{\xi_0}$ such that $RLB_{\xi_0}\approx B_{01}$
\begin{lemma}\label{lem1}
$B_L(p_{val},\xi)=-e\cdot\xi\cdot p_{val}^{\xi}\cdot\log(p_{val})\geq e\cdot\xi\cdot p_{val}^{\xi}>p_{val}^{\xi},$~~~for, $0<p_{val}<e^{-1}$ and $\xi\geq 1$. Note that $B_L(p_{val},1)=B_L(p_{val})$
\end{lemma}
\noindent \begin{proof}

Let $h(p_{val})=-e\cdot\xi\cdot\log(p_{val})$, then $\dfrac{d[h(p_{val})]}{d p_{val}}=-\dfrac{e\cdot\xi}{p_{val}}<0$, thus $h$ is decreasing with minimum at $\xi=e^{-1}$. So, $h(p_{val})\geq h(e^{-1})=e\cdot\xi$ the which implies $B_L(p_{val},\xi)/p_{val}^{\xi}=h(p_{val})\geq e\cdot \xi,$ so $B_L(p_{val},\xi)\geq e\cdot\xi\cdot p_{val}^{\xi}>p_{val}^{\xi}$

\end{proof}

\begin{theorem}
The $\text{RLB}_{\xi}$ is a valid $p$-value, for $\xi\geq 1$, that is, $$P(B_{L}(p,\xi)\leq \alpha|p\sim f(p|\xi))\leq \alpha,\hspace{.2cm}\text{for each}\hspace{.2cm} 0\leq \alpha \leq 1.$$ 
\end{theorem}

\noindent \begin{proof}

First of all it can be seen that $B_L(p,\xi)=-e\cdot\xi\cdot p^{\xi}\cdot\log(p)$ is well defined, since $0\leq B_L(p,\xi)\leq 1.$

Let $\alpha\in [0,1]$, denote for $D_B$ the subset of $R_p$ (range of $p$) such that $$-e\cdot\xi\cdot p^{\xi}\cdot\log(p)\leq \alpha,$$ then  $$(B_L(p,\xi)\leq \alpha)=[-e\cdot\xi\cdot p^{\xi}\cdot\log(p)\leq \alpha]=(p\in D_B)$$
where $(p\in D_B)$ is the event that consists of all the result $x$ such that the point $p(x)\in D_B$. Therefore,

\begin{eqnarray*}
F_B(\alpha)=P(B_L(p,\xi)\leq \alpha|p\sim f(p|\xi))&=&P(-e\cdot\xi\cdot p^{\xi}\cdot\log(p)\leq \alpha|p\sim f(p|\xi))\\
                                             &=&P(p\in D_B|p\sim f(p|\xi))\\
                                             &=& \int_{D_B} f_p(p)dp\\
                                             &=&\int_{0}^{\rho}\xi p^{\xi-1}dp\\
                                             &=&\rho^{\xi}
\end{eqnarray*}

\noindent where $\rho$ is determined such that $$0<\rho<\frac{1}{e}\hspace{.2cm}\text{and}\hspace{.2cm} \alpha=-e\cdot\xi\cdot \rho^{\xi}\cdot\log(\rho)$$ 

as shown in the figure~\ref{fig1} in the case $\xi=1$ 

\begin{figure}
\centering
\includegraphics[scale=.4]{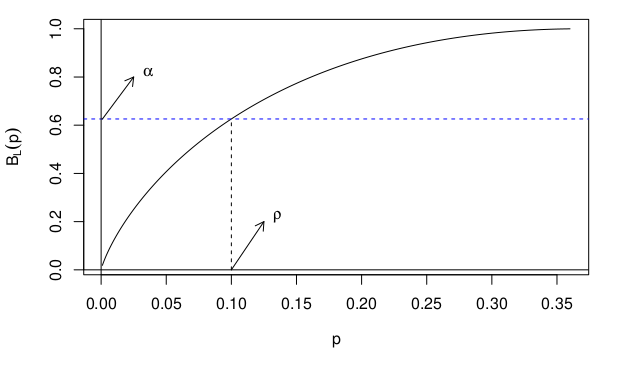}
\caption{Plot of $B_L(p)$ indentifying $\rho$ such that -e$\cdot\rho\cdot\log(\rho)=\alpha$.}\label{fig1}
\end{figure}

now, by Lemma~\ref{lem1} $F_B(\alpha)=\rho^{\xi}<-e\cdot\xi\cdot \rho^{\xi}\cdot\log(\rho)=\alpha$.  
\end{proof}

\section{Adaptive $\alpha$ with Strategy PBIC}

An adaptive $\alpha$ allows us to adapt the statistical significance with the information, but more importantly, it allows us to arrive at equivalent results with a Bayes factor. In \citet{PP2014} this adaptive $\alpha$ based in BIC is presented as:

\begin{equation}\label{eq2}
    \alpha_{n}(q)=\frac{[\chi_{\alpha}^{2}(q)+q\log(n)]^{\frac{q}{2}-1}}{2^{\frac{q}{2}-1}n^{\frac{q}{2}}\Gamma\left(\frac{q}{2}\right)}\times C_{\alpha}
\end{equation}

and in \citet{DPP2022} a version to nested linear models based in PBIC (Prior-based Bayesian Information Criterion, see \citet{Bayarri2019}) is presented as:

\begin{equation}\label{eq3}
\alpha_{(b,n)}(q)=\dfrac{[g_{n,\alpha}(q)+\log(b)+C]^{\frac{q}{2}-1}}{b^{\frac{n-j}{2(n-1)}}\cdot\left(\frac{2(n-1)}{n-j}\right)^{q/2-1}\Gamma\left(\frac{q}{2}\right)}\times \exp\left\{-\frac{n-j}{2(n-1)} \left( g_{n,\alpha}(q)+C \right)\right\},
\end{equation}
where $b=\frac{|\mathbf{X}_j^t\mathbf{X}_j|}{|\mathbf{X}_i^t\mathbf{X}_i|}$  and $\mathbf{X}_i, \mathbf{X}_j$ are design matrix and

\begin{equation*}
    C=2\sum_{m_i=1}^{q_i}\log\frac{(1-e^{-v_{m_i}})}{\sqrt{2}v_{m_i}}-2\sum_{m_j=1}^{q_j}\log\frac{(1-e^{-v_{m_j}})}{\sqrt{2}v_{m_j}},
\end{equation*}
$v_{m_l}=\frac{\hat{\xi}_{m_l}}{[d_{m_l}(1+n^e_{m_l})]}$ with $l=i,j$ corresponding to each model. Here $n^e_{m_l}$, with $l=i,j$, refers to The Effective Sample Size (called TESS) corresponding to that parameter, see (\citet{Bayarri2019}). 

If we adjust (\ref{eq2}) replacing the constant $C_{\alpha}$ with the PBIC strategy the following expression is obtained 

\begin{equation}\label{eq4}
\alpha_{n}(q)=\dfrac{[\chi^2_{\alpha}(q)+q\log(n)+C]^{\frac{q}{2}-1}}{n^{\frac{q}{2}}2^{\frac{q}{2}-1}\Gamma\left(\frac{q}{2}\right)}\times \exp\left\{-\frac{1}{2}\left( \chi_{\alpha}^2(q)+C \right)\right\}.
\end{equation}
Note that this adaptive $\alpha$ is still of BIC structure.
\subsection{Binomial Models}
Consider comparing two binomial models $S_1\sim binomial(n_1,p_1)$ and $S_2\sim binomial(n_2,p_2)$ via the test
$$H_0:p_1=p_2~~vs~~H_1:p_1\neq p_2.$$

Defining $n=n_1+n_2$ and $\hat{p}$ the MLE from $p_1-p_2$, then the equation (\ref{eq4}) is 

\begin{equation}\label{eq5}
\alpha_{n}=\left[\dfrac{2}{n\pi(\chi^2_{\alpha}(1)+\log(n)+C)}\right]^{1/2}\times \exp\left\{-\frac{1}{2}\left( \chi_{\alpha}^2(1)+C \right)\right\},
\end{equation}

here $\chi_{\alpha}^2(1)$ is the quantile $\alpha$ from chi-square with $df=1$, $C=-2\log\dfrac{(1-e^{-v})}{\sqrt{2}v}$, $v=\hat{p}^2/[d(1+n^e)]$, $d=\left(\frac{\sigma_1^2}{n_1}+\frac{\sigma_2^2}{n_2}\right),  n^e=\max\left\{\frac{n_1^2}{\sigma_1^2},\frac{n_2^2}{\sigma_2^2}\right\}d$.\\

The Table~\ref{tab1} shows the behavior $\alpha_n$ when $\alpha=0.5$ and $n_1$ and $n_2$ take different values.

\begin{table}[h]
    \centering
    {\small  \begin{tabular}{|c|c|c|}
	\hline 
	&& Adaptive $\alpha$ via PBIC ($\alpha_n$) \\ 
	\hline 
	$n_1$ & $n_2$ & $n=n_1+n_2$\\ 
	\hline 
	10 & 10 &0.0068\\ 
	25 & 25 &0.0040\\ 
	50 & 50 &0.0027\\
	100 & 50 &0.0021\\
	50 & 100 &0.0021\\ 
	100 & 100 &0.0018\\ 

	\hline 
\end{tabular} }
\caption{Adaptive $\alpha$ via PBIC in equation \ref{eq5} for testing equality of two proportions.}\label{tab1}
\end{table}

\section{Adjusting $RLB_\xi$ with Adaptive $\alpha$}

In this section, we use the equation~(\ref{eq1}) with the adaptive $\alpha$ in equation (\ref{eq3}) and in equation (\ref{eq4}) for obtaining an approximation to an objective Bayes Factor calibrating the $RLB_{\xi_0}$ by The Effective Sample Size (\citet{BBP2014}) and the parameters involved, according to what is established in \citet{art11}.\\ 

Using these ideas, a calibration of (\ref{eq1}) when evaluated in (\ref{eq4}) results in the following Bayes Factor, which has a simple expression.

\begin{equation}\label{eq6}
 B(\alpha,q,n,\xi_0)=-\alpha^{\xi_0}\log(\alpha)\Gamma(q/2)^{\xi_0}n^{\frac{\xi_0q}{2}}\left[\dfrac{2}{\chi^2_\alpha(q)+q\cdot\log(n)+C}\right]^{\frac{\xi_0q}{2}-(\xi_0-1)} . 
 \end{equation}
 
 When it comes to a p-value that is not a pseudo p-value $\xi_0=1$ and the Bayes factor simplifies to
 
 \begin{equation}\label{eq7}
 B(\alpha,q,n)=-\alpha\log(\alpha)\Gamma(q/2)n^{\frac{q}{2}}\left[\dfrac{2}{\chi^2_\alpha(q)+q\cdot\log(n)+C}\right]^{\frac{q}{2}}.  
 \end{equation}

The refined version to linear models, for this calibration is obtained when evaluated in (\ref{eq3}) 

\begin{equation}\label{eq8}
 B(\alpha,q,n,b)=-\alpha\log(\alpha)\Gamma(q/2)b^{\frac{n-j}{2(n-1)}}\left[\dfrac{2(n-1)}{(g_{n,\alpha}(q)+\log(b)+C)(n-j)}\right]^{\frac{q}{2}}  
 \end{equation}
in this case only we consider $\xi_0=1$ since $\alpha$ take value that are not pseudo p-value. 
\subsection{Balanced One Way Anova}
	Suppose we have $k$ groups with $r$ observations each, for a total sample size of $kr$ and let 
$H_0: \mu_1= \cdots = \mu_k=\mu \;\; vs \;\; H_1: \mbox{At least one } \mu_i \mbox{ different}$. Then the design matrices for both models are:

{\footnotesize
\[ \mathbf{X}_1=\left(\begin{array}{c}
1\\ 
1\\ 
\vdots\\ 
1\\
\end{array}\right) \;, \mathbf{X}_k = \left( \begin{matrix}
1 & 0 & \ldots & 0 \\ 
1 & 0 & \ldots & 0 \\ 
\vdots & \vdots & \ldots & \vdots \\ 
1 & 0 & \ldots & 0 \\ 
0 & 1 & \ldots & 0 \\ 
0 & 1 & \ldots & 0 \\ 
\vdots & \vdots & \ldots & \vdots \\ 
0 & 1 & \ldots & 0 \\ 
\vdots & \vdots & \ldots & \vdots \\ 
0 & 0 & \ldots & 1 \\ 
0 & 0 & \ldots & 1 \\ 
\vdots & \vdots & \ldots & \vdots \\ 
0 & 0 & \ldots & 1
\end{matrix}\right) \; , b=\frac{|\mathbf{X}_k^t\mathbf{X}_k|}{|\mathbf{X}_1^t\mathbf{X}_1|}=k^{-1}r^{k-1},\]}

\noindent and the adaptive $\alpha$ for linear model in accordance with what was presented in \citet{DPP2022} is
\begin{equation*}
    \alpha(k,r)=\dfrac{[g_{r,\alpha}(k-1)-\log(k)+(k-1)\log(r)+C]^{\frac{k-3}{2}}}{\left(k^{-1}r^{k-1}\right)^{\frac{r-1}{2(r-1/k)}}\left(\frac{2(r-1/k)}{r-1}\right)^{\frac{k-3}{2}}\Gamma\left(\frac{k-1}{2}\right)}\times \exp\left\{-\frac{r-1}{2(r-1/k)} \left( g_{r,\alpha}(k-1)+C \right)\right\}.
\end{equation*}

Here, the number of replicas $r$ is The Effective Sample Size (TESS).
Therefore, the Bayes factor for this test with respect to equation (\ref{eq8}) is:

\begin{small}
\begin{equation*}
 B(\alpha,k,r)=-\alpha\log(\alpha)\Gamma((k-1)/2)\left(k^{-1}r^{k-1}\right)^{\frac{r-1}{2(r-1/k)}}\left[\dfrac{2(r-1/k)}{(g_{r,\alpha}(k-1)-\log(k)+(k-1)\log(r)+C)(r-1)}\right]^{\frac{k-1}{2}}
\end{equation*}
\end{small}

A very important case arises when $k=2$. For this situation, simplifies to

\begin{equation*}
 B(\alpha,r)=-\alpha\log(\alpha)\left(\dfrac{r}{2}\right)^{\frac{r-1}{2r-1)}}\left[\dfrac{2(r-1)\pi}{(g_{r,\alpha}(1)-\log\left(\dfrac{r}{2}\right)+C)(r-1)}\right]^{\frac{1}{2}}
\end{equation*}




\section{Calibrating $P$-Values}

In this section we will use (\ref{eq6}) and (\ref{eq8}) to determine posterior probabilities for the null hypothesis. Since for any Bayes factor $B_{01}$ $$B_{01}\geq B_L(p,\xi_0) ~~~\text{con}~~~\xi_0\geq 1,~\text{fixed but arbitrary,}$$
a lower bound for the posterior probability of the null hypothesis can be obtained as:

\begin{equation}\label{eq9}
\min P(H_0|Data)=\left[1+\dfrac{1}{B_L(p,\xi_0)}\right]^{-1}.    
\end{equation}

\begin{figure}[h]
    \centering
     \includegraphics[scale=.3]{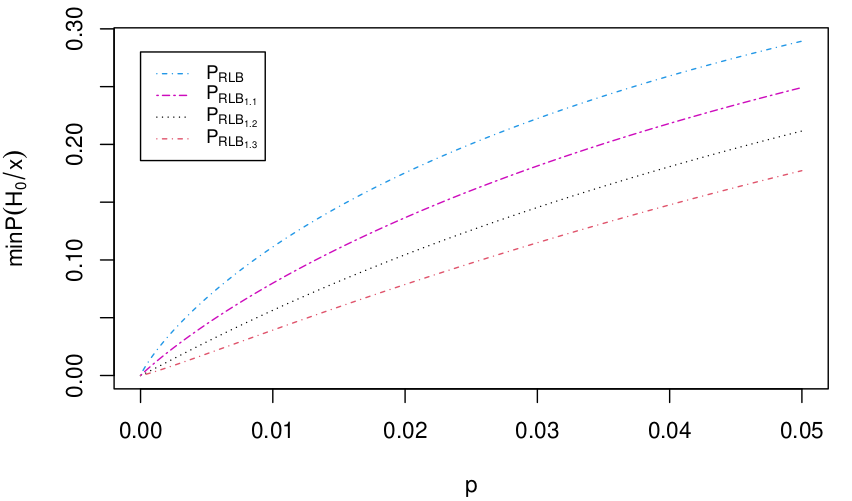}  
    \caption{Lower bound for posterior probability for the null hypothesis $H_0$ for $\xi_0=1, \xi_0=1.1, \xi_0=1.2, \xi_01.3$.}
    \label{fig3}
\end{figure}

The Figure~\ref{fig3} shows these posterior probabilities (called $P_{RLB_{\xi_0}} $) for different values of $\xi_0$

\subsection{Testing Equality of Two Means with Unequal Variances}

Consider comparing two normal means via the test
\begin{equation*}
 H_0:\mu_1=\mu_2 ~~\text{versus}~~ H_1:\mu_1\neq\mu_2,   
\end{equation*} where the associated known variances, $\sigma^2_1$ and $\sigma^2_2$ are not equal. $$\mathbf{Y}=\mathbf{X}\mathbb{\mu}+\mathbb{\epsilon}=\begin{pmatrix}
1&0\\
\vdots&\vdots\\
1& 0\\
0& 1\\
\vdots&\vdots\\
0&1
\end{pmatrix}\begin{pmatrix}
\mu_1\\
\mu_2
\end{pmatrix}+\begin{pmatrix}
\epsilon_{11}\\
\vdots\\
\epsilon_{2n_2}
\end{pmatrix},$$
$$\times \mathcal{\epsilon}\sim N(\mathbf{0},\text{diag}\{\underbrace{\sigma_1^2,...,\sigma_1^2}_{n_1},\underbrace{\sigma_2^2,...,\sigma_2^2\}}_{n_2})$$
Defining $\alpha=(\mu_1+\mu_2)/2$ and $\beta=(\mu_1-\mu_2)/2$ places this in the linear model comparison framework,
$$\mathbf{Y}=\mathbf{B}\binom{\mathbb{\alpha}}{\mathbb{\beta}}+\mathbb{\epsilon}$$
with $$\mathbf{B}=\begin{pmatrix}
1&1\\
\vdots&\vdots\\
1& 1\\
1& -1\\
\vdots&\vdots\\
1&-1
\end{pmatrix}$$
where we are comparing $M_0:\beta=0$ versus $M_1:\beta\neq 0$.\\
So for (\ref{eq8}),
$$C=-2\log\frac{(1-e^{-v})}{\sqrt{2}v}$$
$v=\frac{\hat{\beta}^2}{d(1+n^e)}, d=\left(\frac{\sigma_1^2}{n_1}+\frac{\sigma_2^2}{n_2}\right),  n^e=\max\left\{\frac{n_1^2}{\sigma_1^2},\frac{n_2^2}{\sigma_2^2}\right\}\left(\frac{\sigma_1^2}{n_1}+\frac{\sigma_2^2}{n_2}\right)$.\\
A special case is the standard test of equality of means when $\sigma_1^2=\sigma_2^2=\sigma^2$. Then $$n^e=\min\left\{n_1\left(1+\frac{n_1}{n_2}\right),n_2\left(1+\frac{n_2}{n_1}\right)\right\}.$$

For other hand, considering $\mu=\mu_1-\mu_2$ with $\sigma_1^2=\sigma_2^2=\sigma^2$
\begin{itemize}
    \item $H_0:\mu_1=\mu_2 \longleftrightarrow \mu=0$
     \item $H_0:\mu_1\neq \mu_2 \longleftrightarrow \mu\neq0$
\end{itemize}
Assuming priors
\begin{itemize}
    \item $\mu|\sigma^2,H_1\sim Normal(0,\sigma^2/\tau_0), \tau_0 \in (0,\infty)$
     \item $\pi(\sigma^2)\propto 1/\sigma^2 $ for both $H_0$ and $H_1$.
\end{itemize}
The Bayes factor is: 
\begin{equation*}
 BF_{01}=\left(\dfrac{n+\tau_0}{\tau_0}\right)^{1/2}\left(\dfrac{t^2\frac{\tau_0}{n+\tau_0}+l}{t^2+l}\right)^{\frac{l+1}{2}}   
\end{equation*}
where $$t=\frac{|\bar{\mathbf{Y}}|}{s/\sqrt{n}}$$
t-statistic with degrees of freedom $l=n-1$ and $n=n_1+n_2$ see \citet{Roger2018}.
 
\begin{figure}[h]
    \centering
     \includegraphics[scale=.3]{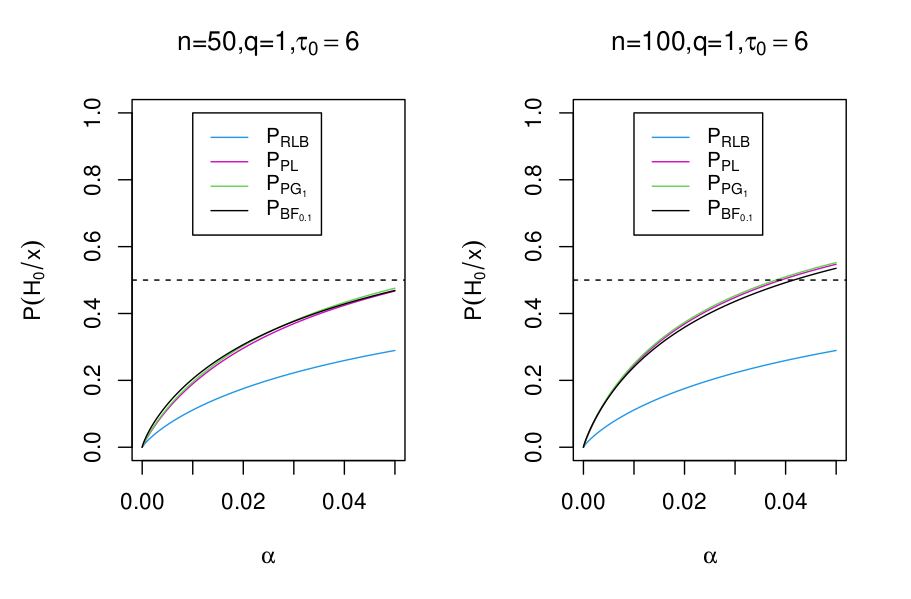}  
    \caption{Posterior probability for the null hypothesis $H_0$ for $n=50$ and $n=100$ using the Bayes factor $RLB_{\xi_0}$ with $\xi_0=1$, the Bayes factor $BF_{01}$, the Bayes factor of the equation (\ref{eq7}) and equation (\ref{eq8}).}
    \label{fig4}
\end{figure}

The Figure \ref{fig4} shows the posterior probability for the null hypothesis $H_0$ when $n=50$ and $n=100$ for the Robust Lower Bound with $\xi_0=1$ (called $P_{RLB}$), the Bayes factor of the equation (\ref{eq8}) (called $P_{PL}$), the Bayes factor of the equation (\ref{eq6}) (called $P_{PG_{\xi_0}}$ with $\xi_0=1$) and for the Bayes factor $BF_{01}$ (called $P_{BF_{0.1}}$). Note that the posterior probability with $BF_{01}$ when $\tau_0=6$ looks very similar to the result obtained using the Bayes factors of the equations (\ref{eq6}) and (\ref{eq8}) . 

\subsection{Fisher's Exact Test}
This is an example where the $p$-value is a pseudo $p$-value (see the example 8.3.30 in \citet{CasBer2001}). Let $S_1$ and $S_2$ be independent observations with $S_1\sim binomial(n_1,p_1)$ and $S_2\sim binomial(n_2,p_2)$. Consider testing $H_0:p_1=p_2$ vs $H_1:p_1\neq p_2$.\\
Under $H_0$, if we let $p$ common value of $p_1=p_2$, the joint pmf of $(S_1,S_2)$ is 
$$f(s_1,s_2|p)=\binom{n_1}{s_1}\binom{n_2}{s_2}p^{s_1+s_2}(1-p)^{n_1+n_2-(s_1+s_2)}$$
and the conditional pseudo $p$-value is
\begin{equation}\label{eq10}
    p(s_1,s_2)=\sum_{j=s_1}^{\min\{n_1,s\}}f(j|s),
\end{equation}
the sum of hypergeometric probabilities, with $s=s_1+s_2$.

It is important to note that in Bayesian tests with point null hypothesis  it is not possible to use continuous prior densities because this distributions (as well as posterior distributions) will grant zero probability to $p=(p_1=p_2)$. A reasonable approximation will be to give $p=(p_1=p_2)$ a positive probability $\pi_0$ and to $p\neq (p_1=p_2)$ the prior distribution $\pi_1g_1(p)$ where $\pi_1=1-\pi_0$ and $g_1$ proper.  One can think of $\pi_0$ as the mass that would be assigned to the real null hypothesis, $H_0:p\in((p_1=p_2)-b,(p_1=p_2)+b)$, if it had not been preferred to approximate by the null point hypothesis. Therefore, if 
$$\pi(p)=\begin{cases} 
    \pi_0 & p=(p_1=p_2) \\
     \pi_1g_1(p) & p\neq (p_1=p_2) \\
      \end{cases}$$
 then 
  \begin{eqnarray*}
      m(s)&=&\int_{\Theta}f(s|p)\pi(p)dp\\
          &=&f(s|(p_1=p_2))\pi_0+\pi_1\int_{p\neq (p_1=p_2)}f(s|p)g_1(p)dp\\
          &=&f(s|(p_1=p_2))\pi_0+(1-\pi_0)m_1(s)\\
      \end{eqnarray*}     
 
\noindent where  $m_1(s)=\displaystyle\int_{p\neq(p_1=p_2)}f(s|p)g_1(p)d p$ is the marginal density of $(S=S_1+S_2)$ with respect to $g_1$.

So, $$\pi((p_1=p_2)|s)=\frac{\pi_0f(s|(p_1=p_2))}{m(s)}$$
thus

\begin{eqnarray*}
\text{odds posterior}&=&\dfrac{\pi((p_1=p_2)|s)}{1-\pi((p_1=p_2)|s)}\\
                          &=&\dfrac{f(s|(p_1=p_2))\pi_0}{m(s)(1-\dfrac{f(s|(p_1=p_2))\pi_0}{m(s)})}\\
                          &=&\dfrac{f(s|(p_1=p_2))\pi_0}{m(s)-f(s|(p_1=p_2))\pi_0}\\
                          &=&\dfrac{f(s|(p_1=p_2))\pi_0}{(1-\pi_0)m_1(s)}\\
                          &=&\dfrac{\pi_0f(s|(p_1=p_2))}{\pi_1m_1(s)}\\
                          &=&\text{odds prior}\cdot \dfrac{f(s|(p_1=p_2))}{m_1(s)}\\
\end{eqnarray*}
     
\noindent and the Bayes Factors is $$B_{01}=\dfrac{f(s|(p_1=p_2))}{m_1(s)}.$$

\begin{figure}[h]
    \centering
     \includegraphics[scale=.3]{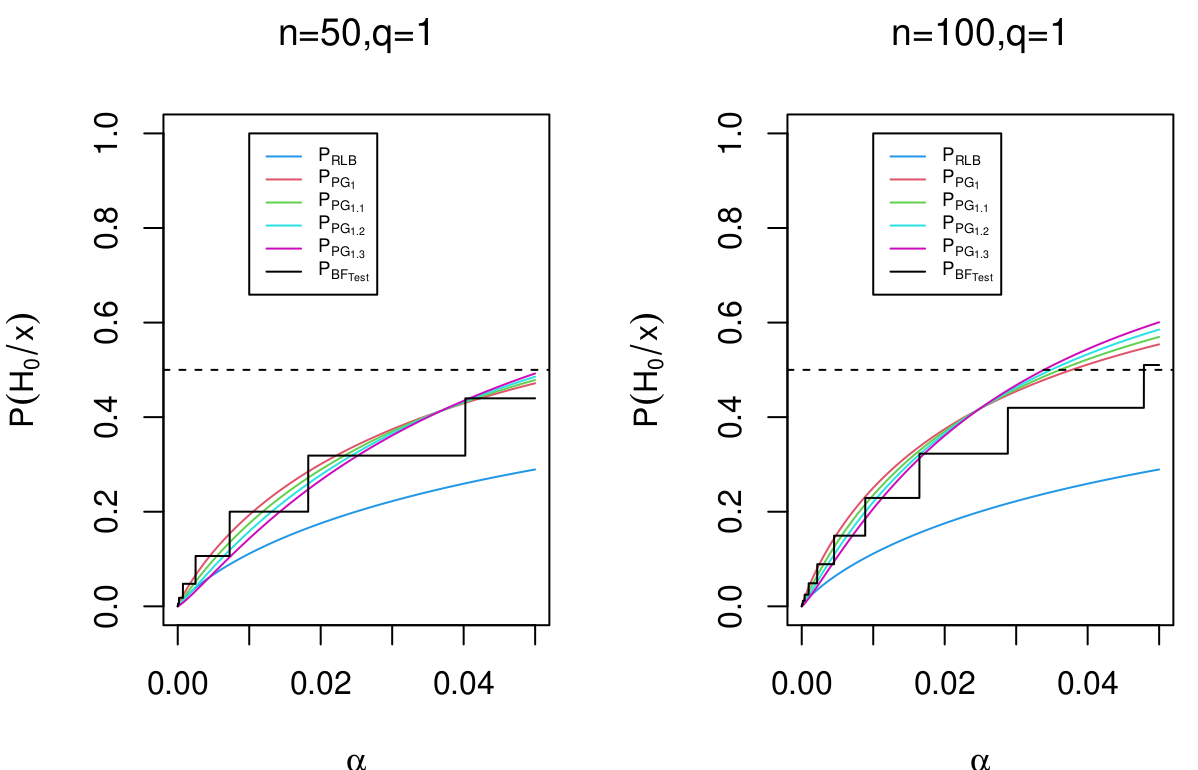}  
    \caption{Posterior probability for the null hypothesis $H_0$ for $n=50$ and $n=100$ using the Bayes factor $RLB_{\xi_0}$ with $\xi_0=1$, the Bayes factor $BF_{Test}$ and the Bayes factor of the equation (\ref{eq6}).}
    \label{fig5}
\end{figure}

Now, if we take $g_1(p)=Beta(a,b)$ such that $E(p)=\dfrac{a}{a+b}=(p_1=p_2)$, then 
\begin{equation*}
   BF_{Test}=\dfrac{B(a,b)}{B(s+a,n_1+n_2-s+b)}p^s(1-p)^{n_1+n_2-s}.
\end{equation*}

the Figure~\ref{fig5} shows the posterior probability for the null hypothesis $H_0$ when $n=n_1+n_2=50$ and $100$ for the Robust Lower Bound with $\xi_0=1$, the Bayes factor of the equation (\ref{eq6}) (called $P_{PG_{\xi_0}}$) and for the Bayes factor $BF_{Test}$ (called $P_{BF_{Test}}$). We can note that all the $P_{PG_{\xi_0}}$  are comparable even though in the case $\xi_0=1$ it is a $p$-value and not a pseudo $p$-value.

\subsection{Linear Regression Models}

Consider comparing two nested linear models $M_3: y_l=\beta_1+\beta_2x_{l2}+\beta_3x_{l3}+\epsilon_l$ with $M_2: y_l=\beta_1+\beta_2x_{l2}+\epsilon_l$ via the test
\begin{equation*}
 H_0:M_2 ~~\text{versus}~~ H_1:M_3,   
\end{equation*}
with $1\leq l\leq n$ and the errors $\epsilon_l$ are assumed to be independent and normally distributed with unknown residual variance $\sigma^2$. According with the equation (\ref{eq3}), in \citet{DPP2022} and in \citet{Bayarri2019} $$b=(n-1)s_{3}^2(1-\rho_{23}^2),$$
where $s_{3}^2$ is the variance $x_{v3}$ and $\rho_{23}$ is the correlation between $x_{v2}$ and $x_{v3}$, and $$C=2\log\frac{(1-e^{-v_{2}})}{\sqrt{2}v_{2}}-2\log\frac{(1-e^{-v_{3}})}{\sqrt{2}v_{3}},$$
where $v_2=\hat{\beta}_2^2/[d_2(1+n^e_2)]$, $d_2=\sigma^2/s^2_{x_{l2}}$, $n^e_2=s^2_{x_{l2}}/\max_i\{(x_{i2}-\bar{x}_2)^2\}$ and $v_3=\hat{\beta}_3^2/[d_3(1+n^e_3)]$, $d_3=\sigma^2(\stackrel{\sim}{X}^t\stackrel{\sim}{X})^{-1}$, $n_3^e=\stackrel{\sim}{X}^t\stackrel{\sim}{X}/\max_i\{|\stackrel{\sim}{X}_i|^2\}$ with $\stackrel{\sim}{X}=(\mathbf{I}_n-X^*(X^{*t}X^*)^{-1}X^*)x_{l3}$
and $X^*=(\mathbf{1}_n|x_{l2})$.

As an example, we analyze a data set taken from \citet{acuna} which can be accessed at \url{http://academic.uprm.edu/eacuna/datos.html}. We want to predict the average mileage per gallon (denoted by \texttt{mpg}) of a set of $n=82$ vehicles using four possible predictor variables: cabin capacity in cubic feet (\texttt{vol}), engine power (\texttt{hp}), maximum speed in miles per hour (\texttt{sp}) and vehicle weight in hundreds of pounds (\texttt{wt}).

Through the Bayes factors in (\ref{eq7}) and (\ref{eq8}) we want to choose the best model to predict the average mileage per gallon by calculating the posterior probability of the null hypothesis of the following test
\begin{center}
$H_0:M_2:$mpg=$\beta_1$+$\beta_2\text{wt}_l$+$\epsilon_l$ vs $H_1:M_3:$mpg=$\beta_1$+$\beta_2\text{wt}_l$+$\beta_3\text{sp}_l$+$\epsilon_l$
\end{center}   

with $\alpha=0.05$, $q=1$, $j=3$, the posterior probabilities for the null hypothesis $H_0$ are: $$P_{PL}=0.9253192, P_{PG_1}=0.7209449$$
where $P_{PL}$ is the posterior probability associated to Bayes factor in equation (\ref{eq8}) and $P_{PG_1}$ is the posterior probability associated to Bayes factor in equation (\ref{eq7}). The use of this posterior probability in both cases will change the inference, since the $p$-value the F test is $p=0.0325$ whose is smaller than $0.05$. 

 \subsubsection{Findley's Counterexample}
 
 Consider the following simple linear model \citep{Findley1991}
 
 $$Y_i=\frac{1}{\sqrt{i}}\cdot\theta+\epsilon_i, ~~\text{where}~\epsilon_i\sim N(0,1),\\
 i=1,2,3,..,n$$
 and we are comparing the models $H_0:\theta=0$ and $H_1:\theta\neq 0$. This  is a Classical and challenging counter example against BIC and the Principle of Parsimony. In \citet{Bayarri2019} it is shown the inconsistency of BIC but the consistency of PBIC in this problem.\\ 
 Here we will show the posterior probabilities of the null hypothesis for this test using the Bayes factors from equations (\ref{eq7}) and (\ref{eq8}) when $n$ grows and $\alpha=0.05$ and $\alpha=0.01$, we will also show the posterior probabilities when $n$ fixed and $0<\alpha<0.05$. For calculations $$C=-2\log\frac{(1-e^{-v})}{\sqrt{2}v}, v=\frac{\hat{\theta}^2}{d(1+n^e)}, d=\left(\sum_{i=1}^{n}\frac{1}{i}\right)^{-1}, n^e=\sum_{i=1}^{n}\frac{1}{i}$$

\begin{figure}[p]
    \centering
     \includegraphics[scale=.25]{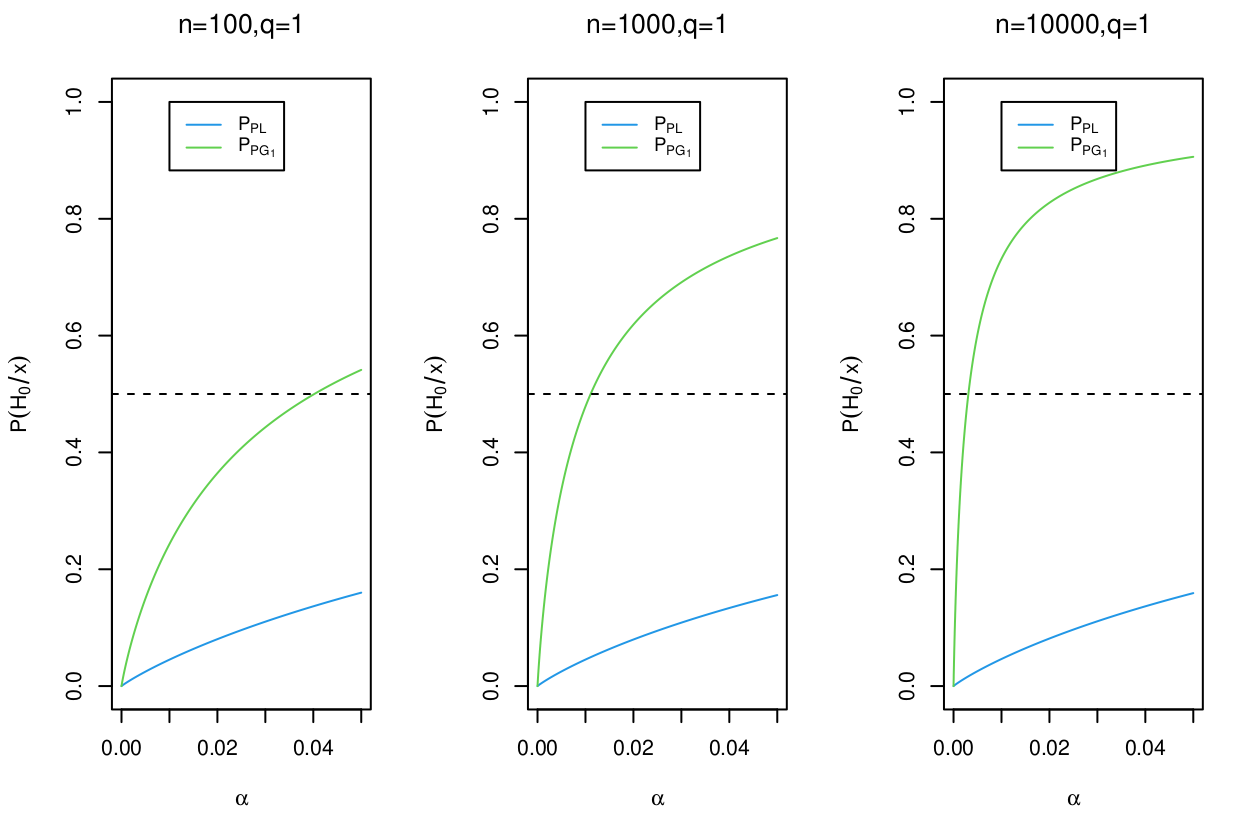}  
    \caption{Posterior probability for the null hypothesis $H_0$ for $n=100$, $n=1000$ and $n=1000$ using the Bayes factor of the equation (\ref{eq7}) and (\ref{eq8}).}
    \label{fig6}
\end{figure}

\begin{figure}[p]
    \centering
     \includegraphics[scale=.25]{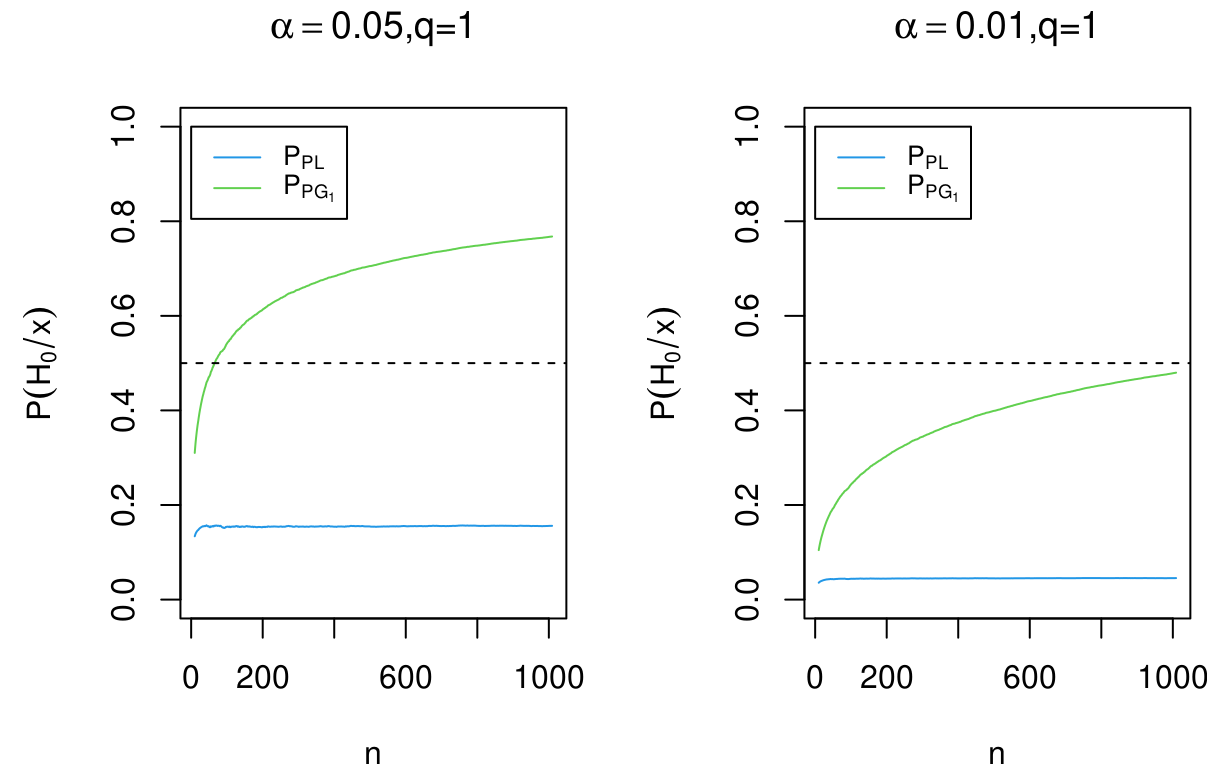}  
    \caption{Posterior probability for the null hypothesis $H_0$ for $\alpha=0.05$ and $\alpha=0.01$ using the Bayes factor of the equation (\ref{eq7}) and (\ref{eq8}) when $n$ grows.}
    \label{fig7}
\end{figure}

The Figure~\ref{fig6} and Figure~\ref{fig7} shows through posterior probability of the null hypothesis $H_0$ the consistency of Bayes factor based in PBIC (equation (\ref{eq8})), and  the inconsistency of Bayes factor based in BIC (equation (\ref{eq7})).

\section{Discussion and Final Comments}

\begin{itemize}
    \item[1.] It will be possible to estimate the appropriate $\xi_0$ that best fits the pseudo p-value in (\ref{eq10})
    \item[2.] The Bayes factors (\ref{eq6}) and (\ref{eq8}) are simple to use and provides results equivalent to the sensitive Bayes factors of hypothesis tests whose p-value may be a pseudo p-value. We hope that this development will give tools to the practice of Statistics.
\end{itemize}

\newpage
\bibliographystyle{chicago}
\newpage
\bibliography{ARLB}

\end{document}